\RequirePackage{ifpdf}
\ifpdf 
\documentclass[pdftex]{sigma}
\else
\documentclass{sigma}
\fi

\begin{document}
\allowdisplaybreaks

\renewcommand{\PaperNumber}{063}

\FirstPageHeading

\ShortArticleName{The Painlev\'e Test and Reducibility to the
Canonical Forms}

\ArticleName{The Painlev\'e Test and Reducibility\\ to the
Canonical Forms for Higher-Dimensional\\ Soliton Equations with
Variable-Coef\/f\/icients}

\Author{Tadashi KOBAYASHI~$^\dag$ and Kouichi TODA~$^\ddag$}
\AuthorNameForHeading{T. Kobayashi and K. Toda}

\Address{$^\dag$~High-Functional Design G, LSI IP Development
Div., ROHM CO., LTD.,\\
$\phantom{^\dag}$~21, Saiin Mizosaki-cho, Ukyo-ku, Kyoto 615-8585,
Japan}
\EmailD{\href{mailto:t-kobayashi@st.pu-toyama.ac.jp}{t-kobayashi@st.pu-toyama.ac.jp}}

\Address{$^\ddag$~Department of Mathematical Physics, Toyama Prefectural University,\\
$\phantom{^\ddag}$~Kurokawa 5180, Imizu, Toyama, 939-0398, Japan}
\EmailD{\href{mailto:kouichi@yukawa.kyoto-u.ac.jp}{kouichi@yukawa.kyoto-u.ac.jp}}

\ArticleDates{Received November 30, 2005, in f\/inal form June 17,
2006; Published online June 30, 2006}

\Abstract{The general KdV equation (gKdV) derived by T.~Chou is
one of the famous $(1 + 1)$ dimensional soliton equations with
variable coef\/f\/icients. It is well-known that the gKdV equation
is integrable. In this paper a higher-dimensional gKdV equation,
which is integrable in the sense of the Painlev\'e test, is
presented. A~transformation that links this equation to the
canonical form of the Calogero--Bogoyavlenskii--Schif\/f equation
is found. Furthermore, the form and similar transformation for the
higher-dimensional modif\/ied gKdV equation are also obtained.}


\Classification{37K10; 35Q53}

\section{Introduction}
Modern theories of nonlinear science have been widely developed
over the last half-century. In particular, nonlinear integrable
systems have attracted much interest among mathematicians and
physicists. One of the reasons for this is the algebraic
solvability of the integrable systems. Apart from their
theoretical importance, they have remarkable applications to many
physical systems such as hydrodynamics, nonlinear optics, plasma
and f\/ield theories and so on \cite{Lamb1980, Akhmediev1997,
Infeld2000}. Generally the notion of the nonlinear integrable
systems \cite{Zakharov1991} is not def\/ined precisely, but rather
is characterized by a number of common features: space-localized
solutions (or {\it solitons}) \cite{Ablowitz1981, Jeffrey1982,
Drazin1989, Ablowitz1991, Hirota2004}, Lax pairs \cite{Lax1968,
Calogero1982, Blaszak1998}, bi-Hamiltonians \cite{Blaszak1998,
Dickey1990}, B\"acklund transformations \cite{Hirota2004,
Rogers2002} and Painlev\'e tests \cite{Ramani1982, Weiss1983,
Gibbon1985, Steeb1989, Ramani1989, Chowdhury1999, Conte2000}.
Moreover, solitons and nonlinear evolution equations are a major
subject in mechanical and engineering sciences as well as
mathematical and physical ones. Among the well-known soliton
equations, the celebrated Korteweg--de Vries (KdV)
\cite{Korteweg1895} and Kadomtsev-Petviashvili (KP)
\cite{Kadomtsev1970} equations possess remarkable properties. For
example, a real ocean is inhomogeneous and the dynamics of
nonlinear waves is very inf\/luenced by refraction, geometric
divergence and so on. The problem of evolution of transverse
perturbations of a wave front is of theoretical and practical
interest. However, f\/inding new integrable systems is an
important but dif\/f\/icult task because of their somewhat
ambiguous def\/inition and undeveloped mathematical background.

The physical phenomena in which many nonlinear integrable
equations with constant coef\/f\/i\-cients arise tend to be very
highly idealized. Therefore, equations with variable
coef\/f\/icients may provide various models for real physical
phenomena, for example, in the propagation of small-amplitude
surface waves, which runs on straits or large channels of slowly
varying depth and width. On one hand, there has been much interest
in the study of generalizations with variable coef\/f\/icients of
nonlinear integrable equations \cite{Calogero1978, Brugarino1980,
Oevel1984, Steeb1985, Chou1987-1, Chou1987-2, Joshi1987,
Hlavaty1988, Brugarino1989, Chan1989, Brugarino1991, Klimek2000,
Gao2001-1, Gao2001-2, Wang2002, Cascava2002, Blanche2004,
Fokas2004, Kobayashi2004, Robbiano2005, Loutsenko2005,
Senthilvelan2005}.

For discovery of new nonlinear integrable systems, many
researchers have mainly investigated ($1 + 1$)-dimensional
nonlinear systems with constant coef\/f\/icients. On the other
hand, there are few research studies to f\/ind nonlinear
integrable systems with variable coef\/f\/icients, since they are
essentially complicated. And their results are still in its early
stages. Analysis of higher-dimensional systems is also an active
topic in nonlinear integrable systems. Since then the study of
integrable nonlinear equations in higher dimensions with variable
coef\/f\/icients has attracted much more attention. So  the
purpose of this paper is to construct  a ($2 + 1$) dimensional
integrable version of the KdV and modif\/ied KdV \cite{Miura1968}
equations with variable coef\/f\/icients.

It is widely known that the Painlev\'e test in the sense of the
Weiss--Tabor--Carnevale (WTC) method \cite{Weiss1983} is a
powerful tool for investigating integrable equations with variable
coef\/f\/icients. We have discussed the following
higher-dimensional 3rd order nonlinear evolution equation with
variable coef\/f\/icients for $u =u(x,z,t)$ \cite{Kobayashi2005-1,
Kobayashi2005-2}:
\begin{gather}
u_t + a(x,z,t)u + b(x,z,t)u_x + c(x,z,t)u_z + d(x,z,t)u u_z  + e(x,z,t)u_x \partial_x^{-1}u_z\nonumber\\
\qquad{}+ f(x,z,t)u_{xxz} + g(x,z,t) = 0, \label{g-cbs}
\end{gather}
where $d(x,z,t) + e(x,z,t) \ne 0$, $f(x,z,t) \ne 0$ and subscripts
with respect to independent variables denote their  partial
derivatives, for example, $u_x = \partial u/\partial x$, $u_{xz} =
\partial^2 u/\partial x \partial z$ etc, and $\partial_x^{-1} u: =
\displaystyle\int^x u(X) d X$. Here $a(x, z, t), b(x,z,t), \ldots,
g(x, z, t)$ are coef\/f\/icient functions of two spatial variables
$x$, $z$ and one temporal variable $t$. We have carried out the
WTC method for equation~(\ref{g-cbs}), and have presented two sets
of the coef\/f\/icient function so that it is shown in next
section. Equations of the form~(\ref{g-cbs}) include one of the
integrable higher-dimensional KdV equations:
\begin{gather}
u_t + u u_z + \frac{1}{2} u_x \partial_x^{-1} u_z + \frac{1}{4}
u_{xxz} = 0,\label{cbseq0}
\end{gather}
which is called the Calogero--Bogoyavlenskii--Schif\/f (CBS)
equation \cite{Calogero1975, Bogoyavlenskii1990, Schiff1992,
Yu1998, Toda1999}. Equation~(\ref{cbseq0}) will be the standard
KdV equation for $u =u(x, t)$:
\begin{gather*}
u_t + \frac{3}{2}u u_x + \frac{1}{4} u_{xxx} = 0
\end{gather*}
by a dimensional reduction $\partial_z = \partial_x$, and the
Ablowitz--Kaup--Newell--Segur (AKNS) equation for $u = u(x, t)$
\cite{Ablowitz1974, Clarkson1994}:
\begin{gather*}
u_t + u u_t + \frac{1}{2}u_x \partial_x^{-1} u_t + \frac{1}{4}
u_{xxt} = 0
\end{gather*}
by another dimensional reduction $\partial_z = \partial_t$. Here
(and hereafter) $\partial_x \equiv \partial/\partial x$,
$\partial_t \equiv \partial/\partial t$ and so on.

This paper is organized as follows. In Section~2, we will review
the process of the WTC method of equation (\ref{g-cbs}) in brief.
In Section 3 and 4 we will consider a general KdV and  a general
modif\/ied KdV equations in ($2 + 1$) dimensions corresponding a
special but interesting case of the integrable higher-dimensional
KdV equation with variable coef\/f\/icients given in Section~2.
Section~5 will be devoted to conclusions.

\section{Painlev\'e test of equation (\ref{g-cbs})}

Let us now brief\/ly review the process of the Painlev\'e test in
the sense of the WTC method for equation (\ref{g-cbs}) put forward
in \cite{Kobayashi2005-1} and then in \cite{Kobayashi2005-2}.

Weiss et al. said in \cite{Weiss1983} that a partial
dif\/ferential equation (PDE) has the Painlev\'e property when the
solutions of the PDE are single-valued around the movable
singularity manifold. They have proposed a technique that
determines whether or not a given PDE is integrable, which we call
the WTC method:

When the singularity manifold is determined by
\begin{gather}
\phi(z_1,\ldots,z_n)=0, \label{series1}
\end{gather}
and $u=u(z_1,\ldots,z_n)$ is a solution of PDE given, then we
assume that
\begin{gather}
u = \sum_{j=0}^{\infty} u_j\phi^{j - \alpha}, \label{series2}
\end{gather}
where $\phi=\phi(z_1,\ldots,z_n)$, $u_j=u_j(z_1,\ldots,z_n)$,
$u_0\ne 0$ ($j = 0,1,2,\ldots$) are analytic functions of~$z_j$ in
a neighborhood of the manifold (\ref{series1}), and $\alpha$ is a
positive integer called the leading order. Substitution of
expansion (\ref{series2}) into the PDE determines the value of
$\alpha$ and def\/ines the recursion relations for $u_j$. When
expansion (\ref{series2}) is correct, the PDE possesses the
Painlev\'e property and is conjectured to be integrable.

Now we show the WTC method for equation (\ref{g-cbs}). For that, a
nonlocal term of equation (\ref{g-cbs}) should be eliminated. We
have a potential form of equation (\ref{g-cbs}) in terms of $U =
U(x, z, t)$:
\begin{gather}
U_{xt} + a(x,z,t)U_x + b(x,z,t)U_{xx} + c(x,z,t)U_{xz} + d(x,z,t)U_x U_{xz} + e(x,z,t)U_{xx} U_z\nonumber\\
\qquad{}+ f(x,z,t)U_{xxxz} + g(x,z,t) = 0\label{pg-cbs},
\end{gather}
when def\/ining $u = U_x$. We are now looking for a solution of
equation (\ref{pg-cbs}) in the Laurent series expansion with $\phi
= \phi (x, z, t)$:
\begin{gather}
U = \sum^{\infty}_{j=0}U_j \phi^{j - \alpha}, \label{expand1}
\end{gather}
where $U_j = U_j (x, z, t)$ are analytic functions in a
neighborhood of $\phi = 0$. In this case, the leading order
($\alpha$) is $1$ and
\begin{gather*}
U_0 = 12 \frac{f(x,z,t)}{d(x,z,t) + e(x,z,t)}\phi_x
\end{gather*}
is given. Then, after substituting of the expansion
(\ref{expand1}) into equation (\ref{pg-cbs}), the recursion
relations for the $U_j$ are presented as follows:
\begin{gather*}
(j+1) (j-1) (j-4) (j-6) f(x,z,t) \phi_x^3 \phi_z U_j = F(U_{j-1},
\ldots, U_0, \phi_t, \phi_x, \phi_z, \ldots),
\end{gather*}
where the explicit dependence on $x$, $z$, $t$ of the right-hand
side comes from that of the coef\/f\/icients. It is found that the
resonances occur at
\begin{gather}
j = - 1, ~1, ~4, ~6\label{res}.
\end{gather}
Then one can check that the numbers of the resonances (\ref{res})
correspond to arbitrary functions~$\phi$, $U_1$, $U_4$ and $U_6$,
though we have omitted the details in this paper. We have
succeeded in f\/inding of two forms of the higher-dimensional KdV
equation with variable coef\/f\/icients given by
\begin{gather}
u_t + \left\{\frac{d'(t)}{d(t)} - \frac{f'(t)}{f(t)} + \frac{4}{3}
\big(\alpha(z,t) - \beta(t) + c_z(z,t)\big)\right\} u
+ \frac{2}{3}x \big(\alpha(z,t) - \beta(t) + c_{z}(z,t)\big) u_x\nonumber \\
\qquad {}+ c(z,t)u_z+ d(t)u u_z + \frac{d(t)}{2}u_x
\partial_x^{-1}u_z + f(t)u_{xxz} + g(z,t) = 0,\label{vc-cbs1}
\end{gather}
and
\begin{gather}
u_t + \big(2 A(z,t) - \eta'(t) \big)u + c(z,t) u_z + \big(A(z,t) x + B(z,t)\big) u_x + d(z,t)u u_z\nonumber\\
\qquad{}+ \frac{d(z,t)}{2}u_x \partial_x^{-1}u_z + \frac{3}{2}
d(z,t) \exp {\eta(t)} u_{xxz} + g(z,t) = 0.\label{vc-cbs2}
\end{gather}
In this paper, $(\cdot)'$ denotes the ordinary derivative with
respect to the independent variable. Apparently, equations
(\ref{vc-cbs1}) and (\ref{vc-cbs2}) are (more general)
higher-dimensional integrable versions of the $(1 + 1)$
dimensional KdV equations with variable coef\/f\/icients appeared
in \cite{Calogero1978, Brugarino1980, Oevel1984, Steeb1985,
Chou1987-1, Chou1987-2, Joshi1987, Hlavaty1988, Brugarino1989,
Chan1989, Wang2002}. It is easy to check by suitable choice of
coef\/f\/icient functions after the dimensional reduction
$\partial_z = \partial_x$.

\section[A general KdV equation in (2 + 1) dimensions]{A general KdV equation in $\boldsymbol{(2 + 1)}$ dimensions}

We consider a special but interesting case of equation
(\ref{vc-cbs1}) in this section.

Setting the following condition of variable coef\/f\/icients:
\begin{gather*}
\alpha(z, t) - \beta(t) = \frac{3}{4} \frac{G^{\prime}(t)}{G(t)},
\quad c(z, t) = 0, \quad d(t) = 1, \quad f(t) = \frac{1}{4}, \quad
g(z, t) = 0,
\end{gather*}
equation (\ref{vc-cbs1}) becomes the following ($2 + 1$)
dimensional equation \cite{Clarkson1997}:
\begin{gather}
u_t + u u_z + \frac{1}{2}u_x \partial_x^{-1}u_z +
\frac{1}{4}u_{xxz} - \frac{G^{\prime}(t)}{G(t)}u - \frac{x}{2}
\frac{G^{\prime}(t)}{G(t)} u_x = 0.\label{gcbseq}
\end{gather}
Here $G(t)$ is an arbitrary function of the temporal variable $t$.
We would like to call equation~(\ref{gcbseq}) a general
Calogero--Bogoyavlenskii--Schif\/f (gCBS) equation. Because
equation~(\ref{gcbseq}) is a~higher-dimensional integrable version
of the general KdV (gKdV) equation \cite{Chou1987-1,
Wang2002}\footnote{Equation (\ref{gcbseq}) is reduced to equation
(\ref{gckdv}) by the dimensional reduction $\partial_z =
\partial_x$.}:
\begin{gather}
u_t + \frac{3}{2}u u_x + \frac{1}{4}u_{xxx} -
\frac{G^{\prime}(t)}{G(t)}u - \frac{x}{2}
\frac{G^{\prime}(t)}{G(t)} u_x = 0.\label{gckdv}
\end{gather}
And using the {\it Lax-pair Generating Technique}
\cite{Kobayashi2005-2, Yu2000, Toda2003}, we obtain a pair of
linear operators associated with equation (\ref{gcbseq}) given by
\begin{gather}
L = \frac{1}{G(t)}\bigr(\partial_x^2 + u \bigr) - \lambda \equiv
 \frac{1}{G(t)}L_{{\rm GKdV}} - \lambda, \label{gcbs-l}\\
T = \partial_z L_{{\rm GKdV}} +
\frac{1}{2}\biggr(\partial_x^{-1}u_z - x
\frac{G'(t)}{G(t)}\biggr)\partial_x - \frac{1}{4}\biggr(u_z -
\frac{G'(t)}{G(t)}\biggr)
 + \partial_t. \label{gcbs-t}
\end{gather}
The compatibility condition\footnote{This is often called the Lax
equation.} of the operators $L$ and $T$\footnote{This pair,
namely, is the Lax pair for equation (\ref{gcbseq}).} is def\/ined
as
\begin{gather}
[L, T] \equiv L T - T L = 0,\label{lax-eq}
\end{gather}
which gives corresponding integrable equations. Notice here that
$\lambda = \lambda(z, t)$ is the spectral parameter and
satisf\/ies the non-isospectral condition \cite{Kobayashi2005-2,
Clarkson1997, Gordoa1999, Estevez2001}:
\begin{gather*}
\lambda_t = \lambda \lambda_z.
\end{gather*}

\begin{figure}[t]
\centerline{\includegraphics[width=7cm]{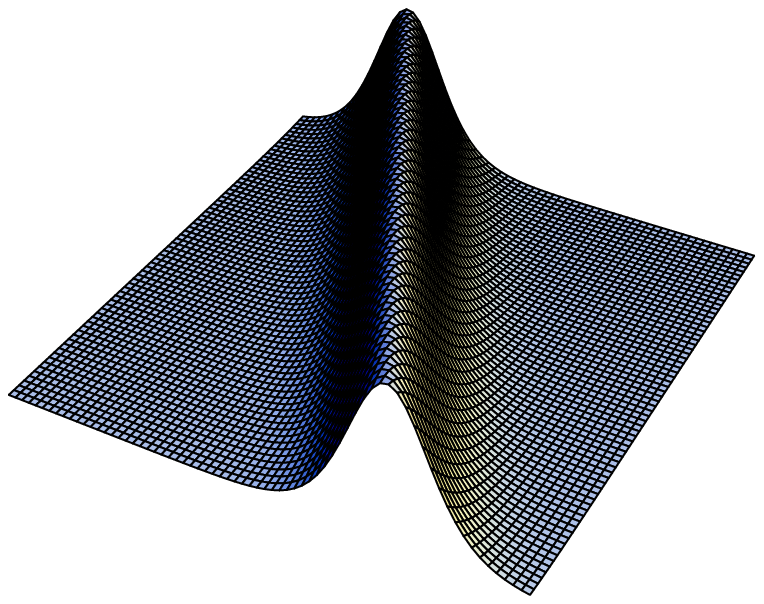} \qquad \quad
\includegraphics[width=7cm]{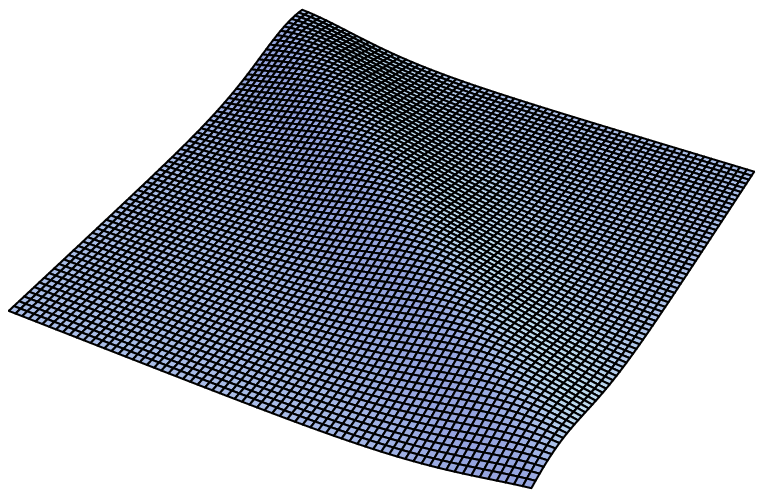}} \caption{One line-soliton
($t$ = 1 $\to$ $t = 5$).} \label{fig:ccbs1}
\end{figure}

\begin{figure}[t]
\centerline{\includegraphics[width=7cm]{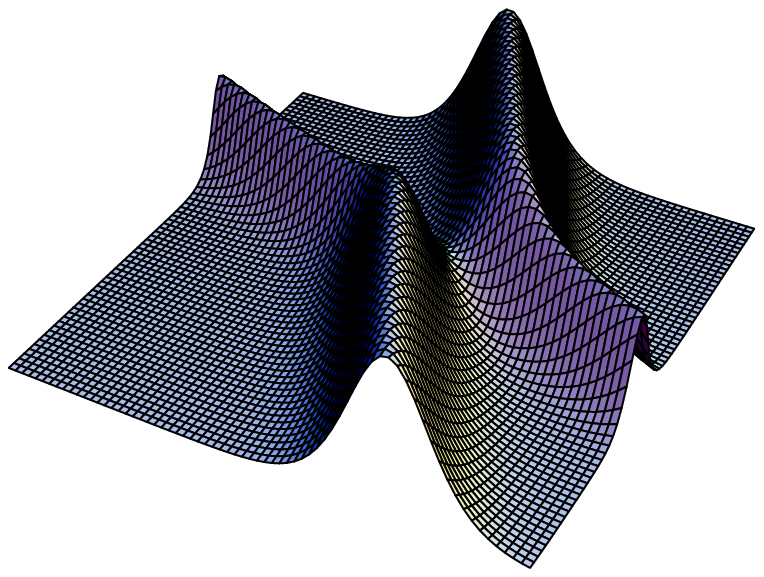}\qquad
\quad\includegraphics[width=7cm]{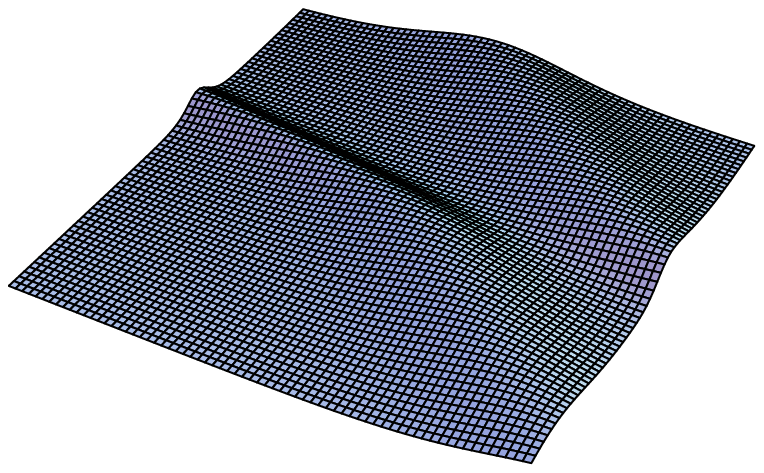}} \caption{Two
line-solitons ($t$ = 1 $\to$ $t = 5$).} \label{fig:ccbs2}
\end{figure}

\begin{figure}[t]
\centerline{\includegraphics[width=7cm]{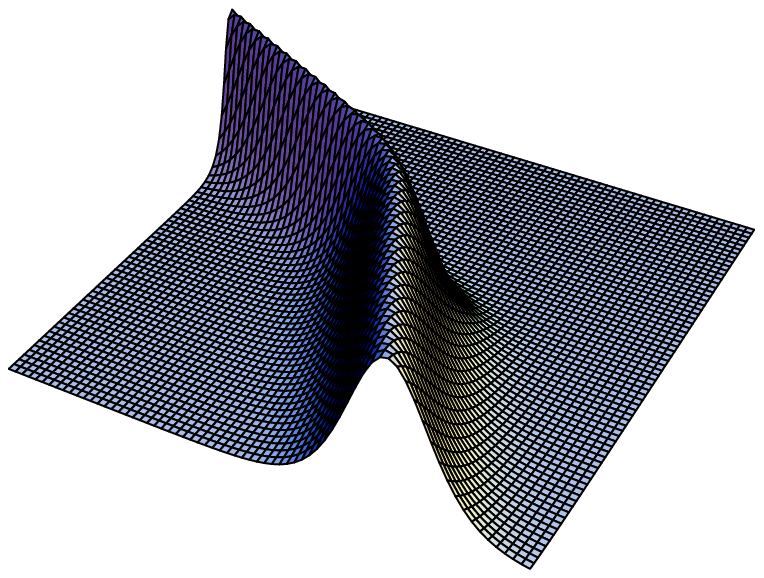}\qquad \quad
\includegraphics[width=7cm]{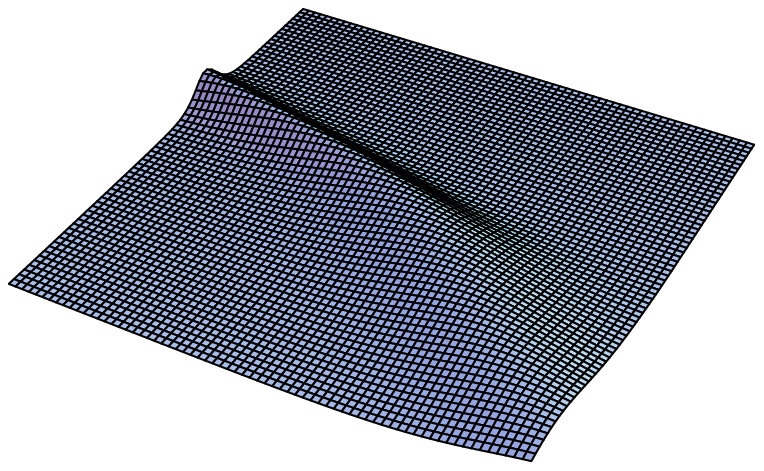}} \caption{V-soliton ($t$ =
1 $\to$ $t = 5$).} \label{fig:ccbsv}
\end{figure}

Let here mention exact solutions of the gCBS equation
(\ref{gcbseq}). As a result of using of the following
transformation:
\begin{gather}
\bar{x} = x \sqrt{G(t)}, \quad \bar{z} = z, \quad \bar{t} =
\partial_t^{-1} G(t), \quad \bar{u}(\bar{x}, \bar{z}, \bar{t}) =
\frac{u(x, z, t)}{G(t)}\label{ch1},
\end{gather}
equation (\ref{gcbseq}) becomes the canonical form of the CBS
equation (\ref{cbseq0}) in terms of $\bar{u} = \bar{u}(\bar{x},
\bar{z}, \bar{t})$ \cite{Bogoyavlenskii1991}:
\begin{gather}
\bar{u}_{\bar{t}} + \bar{u} \bar{u}_{\bar{z}} + \frac{1}{2}
\bar{u}_{\bar{x}} \partial_{\bar{x}}^{-1} \bar{u}_{\bar{z}} +
\frac{1}{4} \bar{u}_{\bar{x}\bar{x}\bar{z}} = 0\label{cbseq1}.
\end{gather}
The $N$ line-soliton solutions with $\tau_{N} = \tau_{N}(\bar{x},
\bar{z}, \bar{t})$ for equation (\ref{cbseq1}) were expressed as
\begin{gather*}
\bar{u} = 2 (\log \tau_N)_{\bar{x}\bar{x}},\\
\tau_N = 1 + \sum^N_{n = 1} \sum_{{}_NC_n} A_{i_1 \cdots i_n}
\exp(\lambda_{i_1} + \cdots + \lambda_{i_n}), \\ 
\lambda_{j} = p_j \bar{x} + q_j \bar{z} + r_j \bar{t} + s_j, \quad j = 1, 2, \ldots, N, \\
r_j = - \frac{p_j^2 q_j}{4},\\
A_{i_1 \cdots i_n} \equiv A_{i_1,i_2} \cdots A_{i_1,i_n} \cdots A_{i_{n - 1},i_n},\\
A_{ij} = \biggr(\frac{p_i - p_j}{p_i + p_j}\biggr)^2,
\end{gather*}
where  the summation ${}_NC_n$ indicates summation over all
possible combinations of $n$ elements taken from $N$, and symbols
$s_j$ always denote arbitrary constants \cite{Yu1998}. So exact
solutions of equation (\ref{gcbseq}) can be presented via the
transformation (\ref{ch1}). We would like to illustrate
line-soliton solutions with $G(t) = 1/t$. Figs.~\ref{fig:ccbs1}
and \ref{fig:ccbs2} are time evolutions of one line-soliton
(with~$p_1$ and~$q_1$) and two line-soliton\footnote{$p_1 \ne
p_2$.} (with $p_1$, $p_2$, $q_1$ and $q_2$) solutions. And a
V-soliton type solution appears on Fig.~\ref{fig:ccbsv} setting
$p_1 = p_2$ in two line-soliton solutions.

A modif\/ied equation corresponding to the gCBS equation
(\ref{gcbseq}) will be given in next section.

\section[A modified general KdV equation in (2 + 1) dimensions]{A modif\/ied general KdV equation in
$\boldsymbol{(2 + 1)}$ dimensions}

We present a higher-dimensional integrable versions of the
modif\/ied gKdV (mgKdV) equation~\cite{Chou1987-2} from the Lax
pair (\ref{gcbs-l}) and (\ref{gcbs-t}) as follows.

Using the {\it Lax-pair Generating Technique}, we obtain a
following Lax pair:
\begin{gather*}
L= \frac{1}{G(t)}\bigr(\partial_x^2 + v \partial_x \bigr) -
\lambda
\equiv \frac{1}{G(t)} L_{{\rm GmKdV}} - \lambda, \\
T= \partial_z L_{{\rm GmKdV}}+ \frac{1}{2}\bigr( \partial_x^{-1}
v_z \bigr)\partial_x^2 + \frac{1}{2} \left(v \partial_x^{-1} v_z -
\frac{1}{4} \partial_x^{-1} \bigr(v^2 \bigr)_z - \frac{1}{2}v_z -
\frac{G'(t)}{G(t)}\right)\partial_x + \partial_{t}.
\end{gather*}
Note here that $\lambda$ satisf\/ies a non-isospectral condition:
\begin{gather*}
\lambda_t = \lambda^2 \lambda_z.
\end{gather*}
Then the Lax equation (\ref{lax-eq}) gives a higher-dimensional
mgKdV equation, or a modif\/ied gCBS equation for $v = v(x, z,
t)$:
\begin{gather}
v_{t}-\frac{1}{4}v^2 v_z -\frac{1}{8} v_x
\partial_x^{-1}\bigr(v^2\bigr)_z + \frac{1}{4}v_{xxz} -
\frac{1}{2} \frac{G'(t)}{G(t)} v - \frac{x}{2}
\frac{G'(t)}{G(t)}v_x = 0.\label{gmcbseq}
\end{gather}
One can also easily check that equation (\ref{gmcbseq}) can be
reduced to the canonical form of the mgKdV equation for $v = v(x,
t)$ \cite{Chou1987-2}:
\begin{gather*}
v_{t}-\frac{3}{8}v^2 v_x + \frac{1}{4}v_{xxx} - \frac{1}{2}
\frac{G'(t)}{G(t)} v - \frac{x}{2} \frac{G'(t)}{G(t)}v_x = 0,
\end{gather*}
by the dimensional reduction $\partial_z = \partial_x$. Via the
transformation as follows:
\begin{gather}
\bar{x} = x \sqrt{G(t)}, \quad \bar{z} = z, \quad \bar{t} =
\partial_t^{-1} G(t), \quad \bar{v}(\bar{x}, \bar{z}, \bar{t}) =
\frac{v(x, z, t)}{\sqrt{G(t)}} \label{ch2},
\end{gather}
equation (\ref{gmcbseq}) becomes the modif\/ied CBS one for
$\bar{v} = \bar{v}(\bar{x}, \bar{z}, \bar{t})$
\cite{{Bogoyavlenskii1991}}:
\begin{gather}
\bar{v}_{\bar{t}} - \frac{1}{4} \bar{v}^2 \bar{v}_{\bar{z}} -
\frac{1}{8} \bar{v}_{\bar{x}} \partial_{\bar{x}}^{-1}
\bigr(\bar{v}^2 \bigr)_{\bar{z}} + \frac{1}{4}
\bar{v}_{\bar{x}\bar{x}\bar{z}} = 0,\label{mcbseq}
\end{gather}
whose $N$ line-soliton solutions were given similarly in
\cite{Yu1998}.

\section{Conclusions}
In mechanical, physical, mathematical and engineering sciences,
nonlinear systems play an important role, especially those with
variable coef\/f\/icients for certain realistic situations. In
this paper we have presented higher-dimensional gKdV and
modif\/ied gKdV equations (\ref{gcbseq}) and (\ref{gmcbseq}),
which are integrable in the sense of the Painlev\'e test. Then we
have found the transformations~(\ref{ch1}) and (\ref{ch2}), which
link equations (\ref{gcbseq}) and (\ref{gmcbseq}) to the canonical
forms (\ref{cbseq1}) and (\ref{mcbseq}), respectively. And also
$N$ line-soliton solutions to equations (\ref{gcbseq}) and
(\ref{gmcbseq}) have been given via the
transformations~(\ref{ch1}) and (\ref{ch2}), respectively.

By applying the (weak) Painlev\'e test\footnote{One can f\/ind the
weak Painlev\'e test for the Camassa--Holm equation in
\cite{Gilson1995, Hone2003, Esteves-Prada2005, Gordoa2006}.}, we
are searching variable-coef\/f\/icient forms of the nonlinear
Schr\"odinger, sine-Gordon, Camassa--Holm and Degasperis--Procesi
equations in ($2 + 1$) dimensions and so on. Here we would like to
give only the form of the ($2 + 1$) dimensional nonlinear
Schr\"odinger equation for $\phi = \phi(x, z, t)$
\cite{Strachan1992, Strachan1993, Jiang1994, Kakei2002}:
\begin{gather*}
i \phi_t + a(x) b(t) \phi_{xz} + b(t) \phi \partial^{- 1}_x
(|\phi|^2)_z + \left\{c(z, t) + \partial_x^{- 1}
\left(\frac{1}{a(x)}\right)\right\} \phi + \frac{1}{2} b(t)
a^{\prime}(x) \phi_z = 0,
\end{gather*}
where $i^2 = -1$. The detail will be reported in
\cite{Kobayashi2005}. We will study relations of nonlinear
integrable equations with variable coef\/f\/icients given in this
paper to realistic situations in mechanical, physical,
mathematical and engineering sciences. Though our program is going
well now, there are still many things worth studying to be seen.

Finally let us mention two forms with variable coef\/f\/icients of
the AKNS equation for $u = u(x, t)$:
\begin{gather}
u_t + \left\{\frac{d'(t)}{d(t)} - \frac{f'(t)}{f(t)} + \frac{4}{3}
\big(\alpha(t) - \beta(t) + c^{\prime}(t)\big)\right\} u
+ \frac{2}{3}x \big(\alpha(t) - \beta(t) + c^{\prime}(t)\big) u_x\nonumber \\
\qquad {}+ c(t) u_t + d(t)u u_t + \frac{d(t)}{2} u_x
\partial_x^{-1} u_t + f(t) u_{xxt} + g(t) = 0,\label{vc-akns1}
\end{gather}
and
\begin{gather}
u_t + \big(2 A(t) - \eta'(t) \big) u + c(t) u_t + \big(A(t) x +
B(t)\big) u_x + d(t) u u_t
+ \frac{d(t)}{2} u_x \partial_x^{-1} u_t\nonumber\\
\qquad{}+ \frac{3}{2} d(t) \exp {\eta(t)} u_{xxt} + g(t) =
0,\label{vc-akns2}
\end{gather}
which are derived from equations (\ref{vc-cbs1}) and
(\ref{vc-cbs2}) by the dimensional reduction $\partial_z =
\partial_t$. Equations~(\ref{vc-akns1}) and (\ref{vc-akns2}) seem
to be new. However it is not certain that equations
(\ref{vc-akns1}) and (\ref{vc-akns2}) are integrable\footnote{One
can check easily that equations (\ref{vc-akns1}) and
(\ref{vc-akns2}) are integrable in the sense of the Painlev\'e
test if $f(t) = d(t)$ for equation (\ref{vc-akns1}) and $\eta$
being a constant for equation (\ref{vc-akns2}).}. We are now
looking for corresponding transformations into the canonical form
of the conventional $(1 + 1)$ dimensional AKNS equation.

\subsection*{Acknowledgements}
Many helpful discussions with Drs.~Y.~Ishimori, A.~Nakamula,
T.~Tsuchida, S.~Tsujimoto, Professors Y.~Nakamura and
P.G.~Est\'evez are acknowledged. One of the authors (K.T.) would
like to thank Professor X.-B.~Hu and his graduate students for
kind hospitality and useful discussions during his stay at the
Chinese Academy of Sciences (Beijing, China) in 2005, where part
of this study has been done. The authors wish to extend their
thanks to anonymous referees for their helpful and critical
comments as this paper took shape.

This work was supported by the First-Bank of Toyama Scholarship
Foundation and in part by Grant-in-Aid for Scientif\/ic Research
(\#15740242) from the Ministry of Education, Culture, Sports,
Science and Technology.

\LastPageEnding

\end{document}